\documentstyle[12pt,psfig]{article}

\input{epsf}
\ifx\epsffile\undefined
\newlength{\epsfysize}
\def\epsffile#1#2#3#4]#5{}
\fi

\catcode`@=11
\long\def\@caption#1[#2]#3{\par\addcontentsline{\csname
  ext@#1\endcsname}{#1}{\protect\numberline{\csname
  the#1\endcsname}{\ignorespaces #2}}\begingroup
    \small
    \@parboxrestore
    \@makecaption{\csname fnum@#1\endcsname}{\ignorespaces #3}\par
  \endgroup}
\catcode`@=12

\newcommand{\met}{{\rlap{\,/}E_T}}
\newcommand{\mplanck}{M_{\rm P}}
\newcommand{\mgut}{M_{\rm GUT}}
\newcommand{\gsim}{\lower.7ex\hbox{$\;\stackrel{\textstyle>}{\sim}\;$}}
\newcommand{\lsim}{\lower.7ex\hbox{$\;\stackrel{\textstyle<}{\sim}\;$}}

\def\ap   #1 #2 #3 #4 {Ann.~Phys.         {\bf  #1}, #2 (#3)#4 }
\def\aplb #1 #2 #3 #4 {Acta Phys.~Pol.    {\bf B#1}, #2 (#3)#4 }
\def\cpc  #1 #2 #3 #4 {Comp.~Phys.~Comm.  {\bf  #1}, #2 (#3)#4 }
\def\epjc #1 #2 #3 #4 {Eur.~Phys.~J.      {\bf C#1}, #2 (#3)#4 }
\def\ijmpa#1 #2 #3 #4 {Int.~J.~Mod.~Phys. {\bf A#1}, #2 (#3)#4 }
\def\jetp #1 #2 #3 #4 {JETP Lett.         {\bf  #1}, #2 (#3)#4 }
\def\mpla #1 #2 #3 #4 {Mod.~Phys.~Lett.   {\bf A#1}, #2 (#3)#4 }
\def\nima #1 #2 #3 #4 {Nucl.~Instrum.~Methods
                                          {\bf A#1}, #2 (#3)#4 }
\def\npb  #1 #2 #3 #4 {Nucl.~Phys.        {\bf B#1}, #2 (#3)#4 }
\def\npps #1 #2 #3 #4 {Nucl.~Phys.~Proc.~Suppl. 
                                         {\bf  #1}, #2 (#3)#4 }
\def\plb  #1 #2 #3 #4 {Phys.~Lett.        {\bf B#1}, #2 (#3)#4 }
\def\prd  #1 #2 #3 #4 {Phys.~Rev.         {\bf D#1}, #2 (#3)#4 }
\def\prl  #1 #2 #3 #4 {Phys.~Rev.~Lett.   {\bf  #1}, #2 (#3)#4 }
\def\ptp  #1 #2 #3 #4 {Prog.~Theor.~Phys. {\bf  #1}, #2 (#3)#4 }
\def\zpc  #1 #2 #3 #4 {Zeit.~Phys.        {\bf C#1}, #2 (#3)#4 }

\renewcommand{\thefootnote}{\fnsymbol{footnote}}
\begin{document}
\begin{flushright}
{\small
FERMILAB-PUB-99/259-T\\
October 7, 1999\\}
\end{flushright}

\vspace{2cm}

\centerline{\bf STOP AND SBOTTOM SEARCHES IN RUN II}
\baselineskip=15pt
\centerline{\bf OF THE FERMILAB TEVATRON}

\baselineskip=35pt

\centerline{\footnotesize REGINA DEMINA}

\baselineskip=20pt

\centerline{\footnotesize\it Department of Physics and Astronomy}
\baselineskip=13pt
\centerline{\footnotesize\it Kansas State University}
\centerline{\footnotesize\it Manhattan, KS 66506, USA}

\baselineskip=32pt

\centerline{\footnotesize JOSEPH D.~LYKKEN and KONSTANTIN T.~MATCHEV}

\baselineskip=20pt

\centerline{\footnotesize\it Theoretical Physics Department}
\baselineskip=13pt
\centerline{\footnotesize\it Fermi National Accelerator Laboratory}
\centerline{\footnotesize\it Batavia, IL 60510, USA}

\baselineskip=32pt

\centerline{\footnotesize ANDREI NOMEROTSKI}

\baselineskip=20pt

\centerline{\footnotesize\it Physics Department}
\baselineskip=13pt
\centerline{\footnotesize\it University of Florida}
\centerline{\footnotesize\it Gainesville, FL 32611, USA}

\vspace{1cm}

\begin{abstract}
We estimate the Tevatron Run II potential for top and bottom
squark searches. We find an impressive reach in several
of the possible discovery channels.
We also study some new channels which may arise in 
non-conventional supersymmetry models. In each case we rely
on a detailed Monte Carlo simulation of the collider
events and the CDF detector performance in Run I.
\end{abstract}


\vfill

\pagebreak
\normalsize\baselineskip=20pt
\setcounter{footnote}{0}
\renewcommand{\thefootnote}{\arabic{footnote}}

\section{Introduction}

For most of the next decade, the Fermilab Tevatron collider
will define the high energy frontier of particle physics. 
The first stage of the Tevatron
collider Run II, scheduled to begin in 2000, 
will deliver at least 2 ${\rm fb}^{-1}$  of integrated luminosity per
experiment at 2.0 TeV center-of-mass energy; this is more than 
10 times the luminosity delivered in previous collider runs at 1.8 TeV. 
Major upgrades of the CDF and D0 detectors are under way.
Among other features, the detectors will have the ability
to trigger on displaced vertices from bottom and charm decays
using a precise microvertex detector.

Along with top physics, searches for supersymmetry (SUSY) are among the 
main priorities for Run II. The Minimal Supersymmetric Standard Model
(MSSM) is an attractive extension
of the Standard Model. On the one hand, it agrees with
precision measurements and cannot be easily ruled out
through indirect searches \cite{EWfits}. On the other hand, it is
theoretically well motivated by various ideas about physics at
very high energy scales -- string theory, supergravity,
grand unification. It also stabilizes the Higgs mass
against radiative corrections, even though it does not provide
the complete answer as to why the electroweak scale is so
much lower than the Planck scale.

Currently, we have no idea which of the supersymmetric particles are
the lightest and as such, more easily accessible at colliders.
With all things being equal, however, the lightest stop 
$\tilde t_1$ is a very good candidate for studying at the Tevatron.
First, it is colored, and so its production cross section is much
larger than that for sleptons, for example. Second, it is feasible
that the stop is the lightest squark. This may be due for example to 
a large mixing angle $\theta_t$ between the superpartners of the
left-handed and the right-handed top quarks,
$\tilde t_L$ and $\tilde t_R$  respectively, 
which form the lightest stop mass eigenstate: 
$\tilde t_1=\tilde t_L\cos\theta_t+\tilde t_R\sin\theta_t$.
Alternatively, the large top Yukawa coupling $\lambda_t$ enters the
renormalization group (RG) equations of the stop soft masses
and tends to reduce them in comparison to the other squarks.
We shall discuss each of these effects 
in more detail in Section~\ref{sec:parspace}.
If the stop is the lightest of all squarks, its
production at the Tevatron will be least suppressed by kinematics.
Yet another motivation to look for a light stop is that
it seems to be preferred for electroweak baryogenesis \cite{EWBG}.

The purpose of this paper is to establish a basis for a
systematic stop search in the upcoming Run II 
at the Tevatron. We shall consider the most
promising stop signatures in a variety of
supersymmetric models. In each case, we shall discuss
under what circumstances the stop can be light,
what is the optimal search strategy, and what is
the Tevatron reach.

In general, the typical SUSY signatures are determined by
the nature of the lightest supersymmetric particle (LSP)
and whether R-parity is conserved or not.
Conservation of R-parity implies that all SUSY decay chains
end up in the LSP, which is stable and leaves the detector.
If the LSP is charge- and color-neutral,
as can often be the case in both supergravity and
gauge-mediated models of supersymmetry breaking,
then a typical SUSY signature is the missing transverse energy $\met$. 
A charged or colored stable LSP could lead to more exotic
signatures (see, for example \cite{light gluino, heavy gluino, Feng}).
R-parity violating interactions \cite{Dreiner} would allow
the LSP to decay into Standard Model particles.
For the purposes of this paper we limit ourselves to
a class of SUSY models where R-parity is conserved and
the LSP is charge and color neutral: i.e. it is either
the lightest neutralino (as in supergravity mediated (SUGRA) models
of SUSY breaking), or the gravitino (as in gauge mediated (GM) models).

The plan of the paper is as follows. In Section~\ref{sec:traditional}
we start with a review of the traditional searches for a light stop,
which were motivated to a large extent by the minimal version (mSUGRA)
of the SUGRA models. Therefore, in Section~\ref{sec:parspace} we first
describe the relevant part of the mSUGRA parameter space which can be
explored via those analyses. Notice, however, that even though we choose
to work within the framework of mSUGRA, our results are valid for a generic
MSSM. In fact, each analysis only makes an assumption about 
the mass ordering of a few supersymmetric particles, and is
not constrained to any particular mechanism of SUSY breaking.
In Section~\ref{sec:exp}
we summarize the assumptions and basic facts from Run I that
were used in the analyses to follow. We then proceed to estimate
the Run II projections for the Tevatron reach in light stop searches
for the following three channels: 
$\tilde t\rightarrow \tilde\chi^+_1 b$ (Section~\ref{sec:chargino}),
$\tilde t\rightarrow \tilde\chi^0_1 c$ (Section~\ref{sec:charm}) and
$\tilde t\rightarrow \tilde \nu \ell^+ b$ or
$\tilde t\rightarrow \tilde \ell^+ \nu b$ (Section~\ref{sec:sneutrino}).
We then depart from the mSUGRA framework,
and consider the possibility of non-universal scalar masses at the
unification scale. This may lead to a situation where the neutralino
LSP is mostly higgsino-like, which will alter the light stop search strategy.
Such a case is studied in Section~\ref{sec:stop to higgsino}, where
we present the reach as a function of the higgsino and stop mass.
As the stop becomes increasingly degenerate with the LSP, the signals
from direct stop production become lost and one has to look for
stops among the decays of other sparticles, e.g. charginos.
In Section~\ref{sec:higgsino to stop} we consider decays of
(possibly higgsino-like) charginos to stops.
We devote Section~\ref{sec:sbottoms} to sbottom searches, since in
many respects the analysis is similar to some of the stop searches
considered before. We delineate the relevant SUGRA parameter
space and present Run II expectations.
Finally, in Section~\ref{sec:GM} we translate our previous results
for the case of light stops in GM models. 
We reserve Section~\ref{sec:conc} for our conclusions.

\section{Traditional Stop Searches}\label{sec:traditional}

In this section we discuss the relevant mSUGRA parameter space
for the traditional stop searches.
We then estimate the Run II sensitivity in these channels.

\subsection{Review of the relevant mSUGRA parameter space}
\label{sec:parspace}

\begin{figure}[t]
\centerline{\psfig{figure=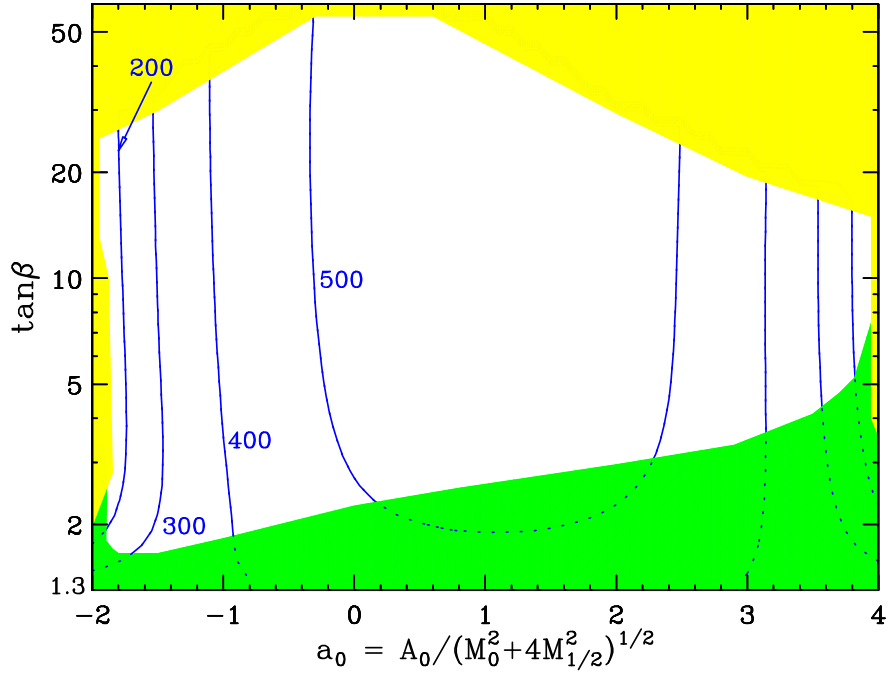,height=3.5in}}
\begin{center}
\parbox{5.5in}{
\caption[] {\small Contours of the light stop mass
$m_{\tilde t}$ (in GeV), versus the mixing parameter
$a_0$ and $\tan\beta$, for $M_{1/2}=300$ GeV,
$M_0=300$ GeV and $\mu>0$.
Inside the light-shaded region there is a scalar
(stop, stau or CP-odd Higgs) which is too light or tachyonic.
The dark-shaded region is theoretically allowed,
but ruled out because of the CP-even Higgs mass limit of
$m_h>95$ GeV.
\label{mstop}}}
\end{center}
\end{figure}

Most of the SUSY searches in the past have been performed within the
SUGRA framework, where supersymmetry breaking, which takes place
in a hidden sector, is communicated to the MSSM fields
through gravitational interactions.
The sparticle spectrum can be calculated in terms of the soft SUSY
breaking parameters, which are in principle free inputs at the Planck scale
$\mplanck$. In the minimal version of the SUGRA models (mSUGRA),
it has become a custom to input the soft masses at the
grand unification (GUT) scale $\mgut$ instead,
and in addition one assumes universality among
the scalar masses, the gaugino masses and the trilinear
Higgs-sfermion-sfermion couplings. The mSUGRA parameter space
therefore consists of only five parameters: a universal scalar mass
$M_0$, a universal gaugino mass $M_{1/2}$, a universal trilinear
scalar coupling term $A_0$, all defined at the GUT scale, as well as
the ratio of the two Higgs vacuum expectation values
$v_u/v_d\equiv\tan\beta$ and the sign of the $\mu$ parameter.
One can readily identify which regions of this
mSUGRA parameter space would be associated with a light stop,
simply by inspection of the stop mass matrix
\begin{equation}
{\cal M}_{\tilde t}=
\left(
\begin{array}{l}
M_{\tilde Q_3}+m_t^2+({1\over2}-{2\over3}\sin^2\theta_W)M_Z^2\cos{2\beta}
\qquad m_t(A_t+\mu\cot\beta)   \\
m_t(A_t+\mu\cot\beta) 
\qquad\qquad\quad
M_{\tilde t_R}+m_t^2+{2\over3}\sin^2\theta_W M_Z^2\cos{2\beta}
\end{array}
\right),
\label{stopmass}
\end{equation}
where $M_{\tilde Q_3}$ ($M_{\tilde t_R}$) is the soft mass parameter
for the left-handed (right-handed) top squark,
$m_t$ is the top quark mass, $M_Z$ is the $Z$-boson mass,
$\theta_W$ is the Weinberg angle and 
$A_t$ is the soft trilinear $\tilde t_L\tilde t_R H_u$ coupling. 
All parameters entering eq.~(\ref{stopmass}) are to be
evaluated at the low-energy scale (near the stop mass).
In the mSUGRA model, $A_t$ is most directly related to its boundary
condition at the GUT scale, $A_0$, while $M_{\tilde Q_3}$
and $M_{\tilde t_R}$ have a dependence on both $M_0$ and $M_{1/2}$.
The $\mu$ parameter is obtained from the condition 
that proper electroweak symmetry breaking reproduces 
the experimentally observed $Z$-boson mass.
Notice the possibility of a large stop mixing:
\begin{equation}
\sin2\theta_t\ \sim \
{2m_t(A_t+\mu\cot\beta)\over m^2_{\tilde t}}
\sim {2m_t A_t\over m^2_{\tilde t}} 
\qquad {\rm (at \ moderate\ to\ large \tan\beta)}.
\label{stopmix}
\end{equation}
If there were no such mixing, the stop masses would be roughly
equal to the corresponding soft mass parameters
$M_{\tilde Q_3}$ and $M_{\tilde t_R}$.
The mixing, however, further reduces the smaller of the two
mass eigenstates \cite{stop mass eigenvalue}.
(This effect is negligible for the first two
generation squarks, because the corresponding quark mass
entering eq.~(\ref{stopmix}) is very small.) 
Since $A_t$ is directly related to its boundary condition
$A_0$ at the GUT scale, one would expect that mSUGRA models
with large values of $|A_0|$ would have a light stop
in their spectrum. As it turns out, a better parameter
for light stop discussions is the dimensionless ratio
$a_0\equiv A_0/\sqrt{M_0^2+4M^2_{1/2}}\simeq A_t/m_{\tilde t}$,
which can be easily understood from eqs.~(\ref{stopmass})
and (\ref{stopmix}). Increasing either $M_0$ or $M_{1/2}$
would increase the diagonal entries in the stop mass matrix,
and make the stop heavier, while $A_0$ controls the size of $A_t$
at the weak scale. In Fig.~\ref{mstop} we show
contours of the lightest stop mass (in GeV) in the $(a_0,\tan\beta)$
plane, for $M_{1/2}=300$ GeV, $M_0=300$ GeV and $\mu>0$. 
Inside the light-shaded region there is a scalar
(stop, stau or CP-odd Higgs) which is too light or tachyonic.
The dark-shaded region is theoretically allowed,
but ruled out because of the CP-even Higgs mass limit of
$m_h>95$ GeV. In calculating the
light stop mass, we have included the full one-loop corrections to
the stop mass matrix \cite{PBMZ}. This is needed in order to reduce
the scale dependence of $m_{\tilde t_1}$, which is known to
be significant, especially in such cases of large stop mixing.

We see from Fig.~\ref{mstop} that a light stop is associated 
with a limited range of rather large values of $|a_0|$,
where the stop mixing becomes very large. We have checked,
by scanning the mSUGRA parameter space, that if $\tan\beta\gsim 2.0$,
values of $a_0$ between 0 and 2 would never lead to stop masses
below 200 GeV. 

\begin{figure}[t]
\centerline{\psfig{figure=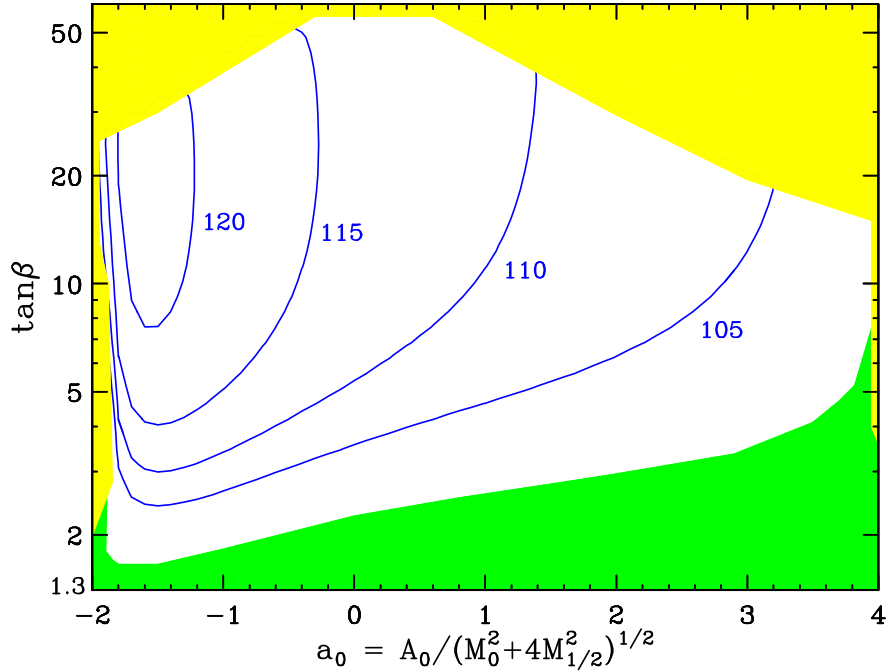,height=3.5in}}
\begin{center}
\parbox{5.5in}{
\caption[] {\small Contours of the light Higgs mass $m_h$ (in GeV)
for the same parameters as in Fig.~\ref{mstop}.
\label{mh}}}
\end{center}
\end{figure}

Another possibility to have a light stop
is that the soft mass parameters
$M_{\tilde Q_3}$ and $M_{\tilde t_R}$ entering eq.~(\ref{stopmass})
are small by themselves.
This may be due to the RG evolution down from very high scales,
which tends to order the squark soft masses in an inverse hierarchy
with respect to their Yukawa couplings.
Since the top Yukawa coupling $\lambda_t$ is so large,
the stops ``feel'' this RGE effect to a larger
extent than the other squarks.
As a result, the stop soft masses $M_{Q_3}$
and $M_{\tilde t_R}$ are typically smaller than
the soft mass parameters of the other squarks.
This effect is strongest when the top Yukawa coupling is the largest,
i.e. at small values of $\tan\beta$. Indeed, Fig.~\ref{mstop}
exhibits a region at rather small values of $\tan\beta$,
where one may find a light stop for a very wide range of the
$A_0$ parameter.
Very low values of $\tan\beta$, however, are excluded
both experimentally and theoretically. In the MSSM,
the lightest Higgs mass $m_h$ is correlated with $\tan\beta$
and current bounds on $m_h$ from LEP all but exclude values
of $\tan\beta\lsim 2.0$. In Fig.~\ref{mh} we show contours of
the light Higgs mass $m_h$, for the same parameters as in
Fig.~\ref{mstop}.
We see that the current Higgs bound already rules out the region where
we get a light stop because of the large Yukawa RGE effects.
This constraint may be relaxed in models with a non-minimal
Higgs sector, like the NMSSM \cite{NMSSM}. In that case,
there is still a theoretical lower bound on $\tan\beta\lsim 1.3-1.5$,
which is due to the requirement that $\lambda_t$
remains perturbative when extrapolated up to very high scales,
e.g. $\mplanck$. What is more, from Fig.~\ref{mh} we also see
that the direct Higgs searches in Run II \cite{Higgs searches}
also eat away from the light stop parameter space. 
If the Higgs escapes detection after the first stage of Run II 
(2 ${\rm fb}^{-1}$ per detector), the expected $m_h$ limit 
with (without) a neural network improvement of the analysis
will be $m_h>120$ (105) GeV \cite{Higgs searches}.
In that case, the mSUGRA light stop parameter space at
$a_0>-1$ ($a_0>3$) will also be ruled out. However, the light stop
region at $A_0<0$ may actually be more effectively probed
via stop searches rather than via the Higgs search.

\begin{figure}[t]
\centerline{\psfig{figure=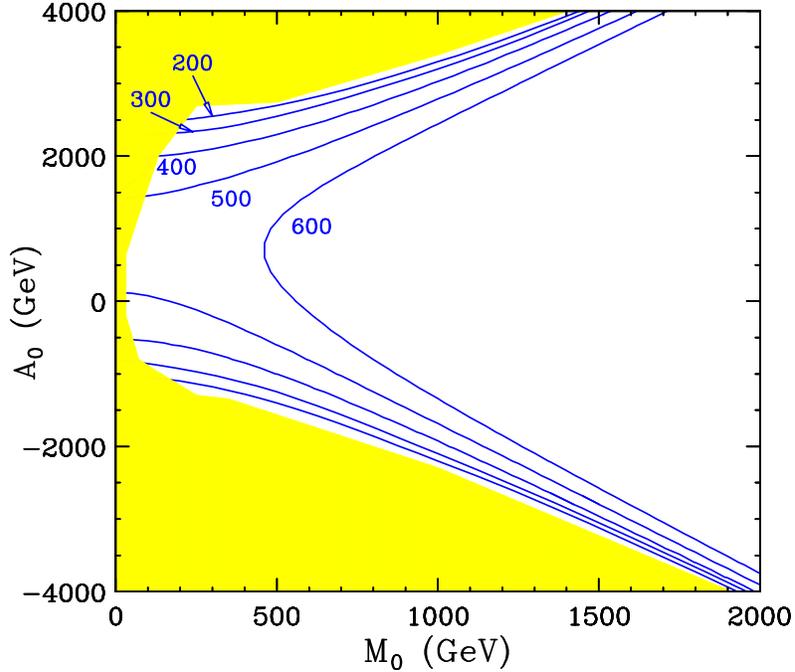,height=3.5in}}
\begin{center}
\parbox{5.5in}{
\caption[] {\small Contours of the light stop mass $M_h$ (in GeV)
in the $M_0-A_0$ plane, for fixed $M_{1/2}=300$ GeV,
$\tan\beta=10$ and $\mu>0$.
\label{m0_a0_mstop}}}
\end{center}
\end{figure}

It is straightforward to check that the (very light) stop mass has
no significant dependence on the other mSUGRA parameters.
Our mSUGRA scan reveals that there exist very
light stop solutions for almost any values of
$M_0$ and $M_{1/2}$! The reason is that there always exist
large enough values of $|A_0|$, which can increase the amount of stop mixing
and yield an arbitrarily light stop in the spectrum.
This is illustrated in Fig.~\ref{m0_a0_mstop},
where we show contours of the light stop mass in the $M_0-A_0$ plane,
for fixed $M_{1/2}=300$ GeV, $\tan\beta=10$ and $\mu>0$.
This value of $M_{1/2}$ is just beyond the
Tevatron reach for chargino-neutralino searches in the trilepton
channel \cite{trileptons}. We then see that for
arbitrarily large $M_0$, i.e. heavy scalars, 
there is still a (rather limited) range of $A_0$, where
of all SUSY searches in Run II,
only the light stop searches have a chance of being successful.
Of course, the branching ratios of the stop decays do depend
on both $M_0$ and $M_{1/2}$, and this will determine the
degree of applicability of each of the channels that we are 
considering below. We shall comment on the relevant range
for $M_0$ and $M_{1/2}$ in each case as we go along.
Also notice that the production cross section for stops is
almost independent of any of the MSSM parameters, and is uniquely
determined by the stop mass.

We should point out that all stop search analyses below
are pretty much model independent and have only a few mild
assumptions about the sparticle spectrum. The motivation
for having a light stop in the spectrum, on the other hand,
is very model-dependent. For example, in mSUGRA very large
values of $|a_0|$ violate the naturalness criterion
\cite{naturalness}, implying that the odds for a light stop
in mSUGRA are rather small. But the odds for mSUGRA being the
correct model of SUSY breaking are probably pretty small too.
So we adopt the stance that we should leave any theoretical
prejudice behind and look for the stop in every conceivable
channel, until the first experimental SUSY data are in.

\subsection{Experimental assumptions}\label{sec:exp}

To determine the mass reach for various search modes in Run~II we compare 
the number of expected signal 
events to a variation of the number of background events (assumed  
to be purely poissonian). A mass region is excluded if the signal is 
larger than 3 standard deviations of the background.

The number of events is calculated as a product of 
the integrated luminosity, production cross section and efficiency after final
cuts.

The integrated luminosities of 2 fb$^{-1}$, 4 fb$^{-1}$ and 
20 fb$^{-1}$ were used as expectations for different stages of Run~II.

At the Tevatron, third generation scalar quarks are expected to be produced in
pairs via gluon-gluon fusion and quark-antiquark annihilation, and, 
consequently, the leading-order production cross section depends only on 
their masses.
The next-to-leading order corrections increase the cross section and 
introduce a weak dependence on other masses and parameters ($\sim$ 1\%) 
\cite{stop xsec}. Signal events were modeled using the PYTHIA generator 
\cite{PYTHIA}. 

In the following we assume that the increase in the production cross section 
due to variation of the center-of-mass energy from 1.8 to 2.0 TeV will 
be approximately 40\% for the squark pair production. Background cross 
sections were also scaled with factors determined from the corresponding
Monte Carlo generators \cite{VECBOS} (20\% for the $W$/$Z$ production and QCD).

Our Run~II efficiency estimates are based on the performance of the CDF 
detector in Run~I using the full detector simulation. 
When possible we extrapolated the 
results of existing CDF searches in corresponding channels. 

CDF is a general purpose detector described in detail 
elsewhere \cite{CDF}. The innermost part of CDF, a silicon vertex detector 
(SVX), allows a precise measurement of a track's impact parameter with respect
to the primary vertex in the plane transverse to the beam direction. The 
momenta of the charged particles are measured in the central drift chamber
which is located inside a 1.4 Tesla solenoidal magnet. Outside the drift 
chamber there is a calorimeter, which is organized into electromagnetic and 
hadronic components, with projective towers covering the pseudo-rapidity range 
$|\eta|<4.2$. The muon system is located outside the calorimeter and covers 
the range in $|\eta|<1$.

Jet energies are calculated using the calorimeter energy deposition within 
a cone in $\eta-\phi$ space, where $\phi$ is the angle in the plane normal to 
the beam direction.   
The missing transverse energy is defined as the energy imbalance in the 
directions transverse to the beam direction using the energy deposited in 
calorimeter towers with $|\eta|<3.7$. 
A lepton is identified by either hits in a muon chamber or a cluster of energy 
in the electromagnetic calorimeter, and an associated track in the central 
tracking chamber. 

Two heavy flavor tagging algorithms use the SVX information to tag charm 
and bottom quark jets, or c-jets and b-jets. 
In the Jet Probability (JP) algorithm
the probability that the track comes from the primary vertex is determined 
taking into account the impact parameter resolution. This probability is 
smaller for heavy flavor decay products because of their considerable lifetime.
The track probabilities for tracks associated to a jet are 
combined into JP \cite{JP}. We associate 
tracks to a jet by requiring that the track is within a cone of 0.4 in 
$\eta-\phi$ space around the jet axis. Distribution of JP is flat 
by construction for light quark jets, originating from the primary vertex, 
and peaks at zero for heavy quarks. Since the JP is a continuous variable,
the tagging can be optimized both for charm and bottom jets.
In another heavy flavor tagging algorithm, SECVTX, 
a jet is identified as a bottom quark candidate if 
its decay point is displaced from the primary vertex \cite{SECVTX}. 
This algorithm is not very efficient for the charm tagging.

In calculating Run~II efficiencies we scaled by a factor of two 
the Run~I efficiency for heavy flavor tagging. 
This factor accounts for the increase 
in the SVX geometrical acceptance due to the larger SVX length. 
Conservatively, we did not assume any other improvements in the detector 
performance such as, for example, a better coverage for the 
lepton identification. 
The description of the upgraded CDF-II detector can be found in 
\cite{CDF2}.

\subsection{Reach in the $bj\ell\met$ channel}\label{sec:chargino}

If the chargino (either gaugino-like or higgsino-like)
is lighter than the light stop, the dominant decay of the
stop is $\tilde t\rightarrow b \tilde\chi^+_1$
\cite{HK,Bartl91,BDGGT,BST,stop to chargino}.
If there are no sfermions (squarks or sleptons) lighter than the chargino,
the latter decays to a real or virtual $W$ and the lightest neutralino.
In this case stop decays produce top-like signatures: 2$W$'s and 2 b-jets.
The only differences are kinematical : higher $\met$ due to the massive 
neutralinos, different jet spectra and angular distributions. 

There are two possible search strategies for this decay mode based on different 
signatures : $b\ell\,j\met$ and $\ell^+\ell^- j \met$. 
We note that in the case of the 
dilepton signature one pays a price of the low lepton branching ratio twice.
Here we present a sensitivity study based on the $b\ell j \met$ signature.
We select events with an isolated electron or muon with $p_{T}>10$ GeV/c 
passing lepton identification cuts, and
at least two jets, one with $E_T> 12$ GeV and the second with $E_T> 8$ GeV. 
At least one of the jets is b-tagged with the SECVTX algorithm.
To decrease Drell-Yan and $Z^0$ background we removed events with two 
isolated, opposite sign leptons.
For the reduction of QCD background we also required $\met > 25$ GeV
and $\Delta\phi(\met$~-~nearest~jet)~$> 0.5$.

The main remaining backgrounds for this search are from the
$W+{\rm jets}$ and top production. Table~\ref{backgrounds} lists 
the relative contributions of different backgrounds after all cuts
for this channel together with two other experimental stop signatures to be 
considered later.
In the table we quote the $W(\rightarrow \tau\nu)+$jets contribution
separately
because of possible hadronic decays of $\tau$ which mimic a jet. The last 
line shows the total background cross section after final cuts. 

\begin{table}[t!]
\centering
\renewcommand{\arraystretch}{1.5}
\begin{tabular}{||c||c|c|c||}
\hline\hline
Background & \multicolumn{3}{c||}{Stop signature} \\
\cline{2-4}
process    & $b\ell\ j \met$   
             & $cc\met$   
               & $\ell^+\ell^- j \met$ 
                       \\  \hline\hline      
$W(\rightarrow e(\mu)\nu)+$jets  & 52\%
             & 1\% 
              & -
                       \\ \hline
$W(\rightarrow \tau\nu)+$jets   & 3\%
             & 52\%
              & -
                      \\ \hline
Drell-Yan,$Z+$jets   & 2\%
             & 8\%
              & 26\%
                       \\ \hline
$ WW/WZ/ZZ$ & - 
              & 3\%
               & 13\%
                        \\ \hline
$t\bar{t}$ & 21\%
             & 5\%
              & 21\%
                       \\ \hline
QCD (includes $b\bar{b}$)        & 20\%
             & 23\%
              & 41\%
                      \\ \hline\hline
Total cross section, fb        & 980
             & 160
              & 50
                      \\ \hline\hline
\end{tabular}
\parbox{5.5in}{
\caption{ Relative contribution of various backgrounds for the stop searches 
after final cuts and the total background cross sections.
\label{backgrounds}}}
\end{table}

\begin{figure}[t]
\centerline{\psfig{figure=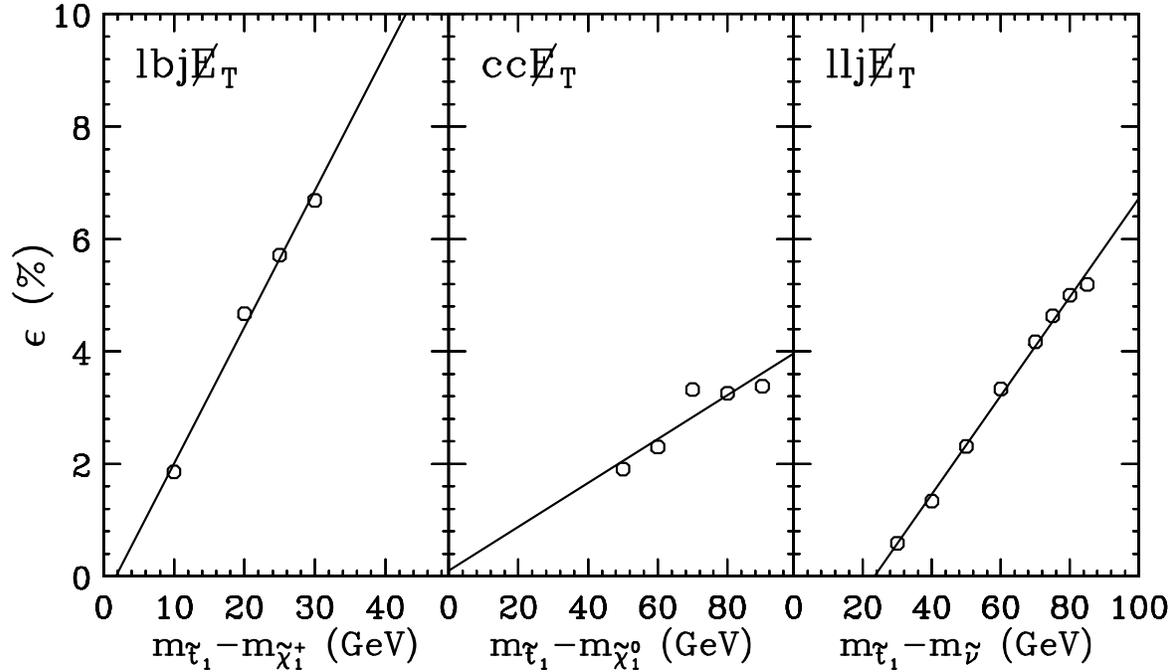,height=3.5in}}
\begin{center}
\parbox{5.5in}{
\caption[] {\small Typical values of efficiencies after final cuts for the
stop signatures considered in the text.
\label{eff}}}
\end{center}
\end{figure}

Typical values of efficiencies after final cuts for this and other
(considered below) stop signatures are shown in Fig.~\ref{eff}.
The efficiencies are plotted as a function of
the mass difference between the stop and its supersymmetric decay product.
The mass difference effectively determines the kinematical properties of the 
reaction and, therefore, the efficiency. Straight line fits show the
efficiency parametrizations used for our estimates.

\begin{figure}[t]
\epsfysize=3.5in
\epsffile[-130 155 170 635]{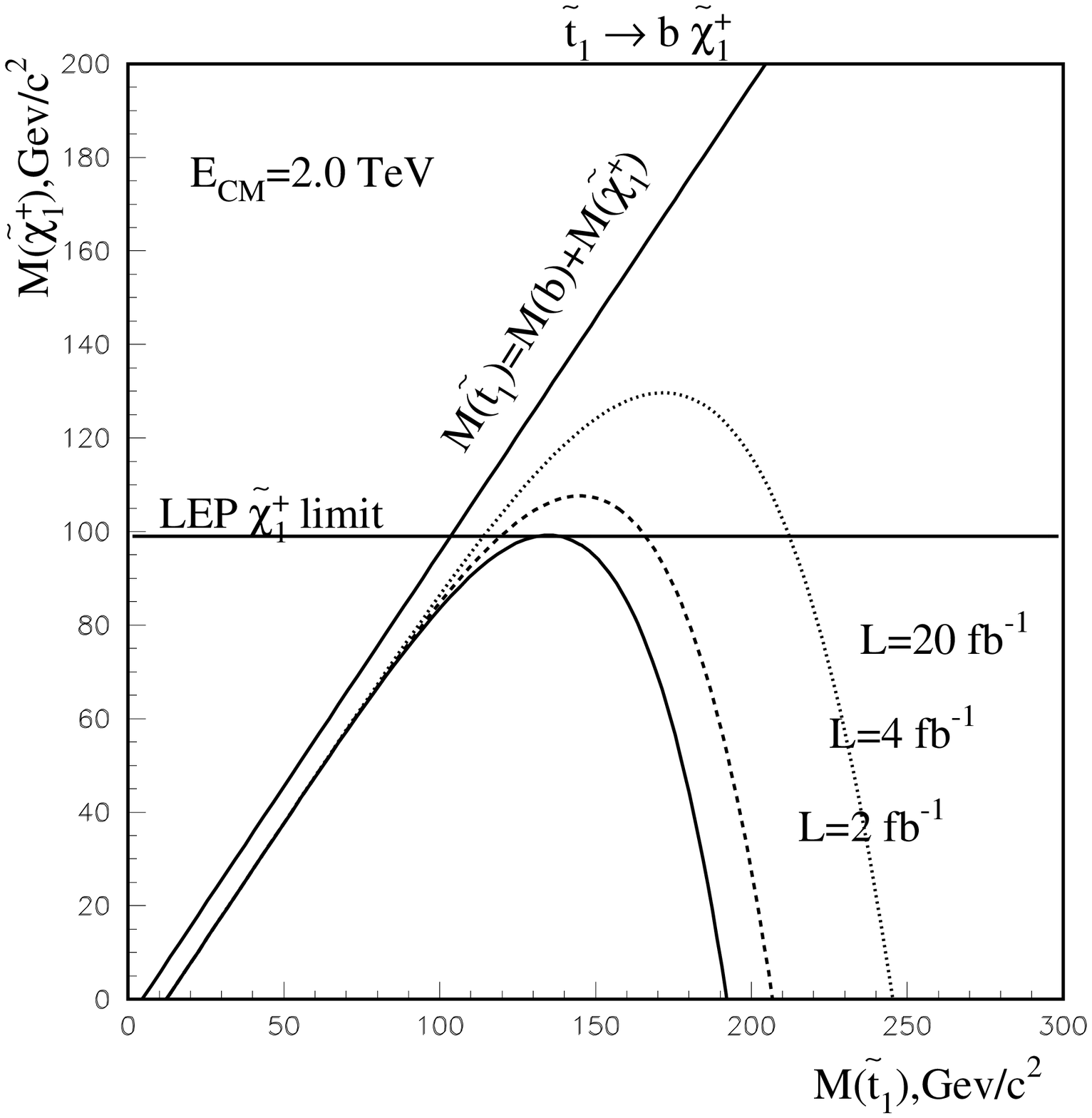}
\begin{center}
\parbox{5.5in}{
\caption[]
{\small Sensitivity of the light stop search in 
the $b\ell j \met$ channel for several integrated luminosities.
The analysis assumes 100\% branching ratio for
$\tilde t\rightarrow b \tilde\chi^+_1$ and $W$-like
branching ratios for the chargino decays.
\label{sens_bchar}}}
\end{center}
\end{figure}

We show the reach of Run II in this channel in Figure~\ref{sens_bchar}.
Even with 2 ${\rm fb}^{-1}$ of data we will be sensitive to stop masses
up to $m_t$. The sensitivity vanishes when we approach the kinematic
limit for this channel, because the b-jets become too soft.
We see also that the projected reach for 2 ${\rm fb}^{-1}$ 
completely overlaps with the region of expected
sensitivity for chargino searches at LEP, but with 
higher luminosity the Tevatron will be able to extend those
bounds.

Note that the leptonic branching ratio of the chargino can be
significantly increased, if the (left-handed) sleptons are relatively
light, which happens when $M_0\lsim M_{1/2}$. This would lead to
a much larger Tevatron reach in this channel \cite{light slepton}.
In the mSUGRA model we find that for $M_0 \le 0.55 M_{1/2}$
the electron and muon sneutrinos are lighter than the
chargino, which would increase the chargino
branching ratio to leptons to almost 100\%.
In this case the dilepton signature, 
discussed in detail in \ref{sec:sneutrino},
may give a better sensitivity.

When the sneutrinos are light, the leptonic branching ratios of gauginos
are high and this region of parameter space can also be effectively
probed via searches for chargino/neutralino production in
the clean trilepton channel \cite{trileptons}.

A possible improvement in the analysis would be a requirement of an
additional b-jet, given a much higher integrated luminosity
in Run~II. This cannot be afforded for Run~I analyses because the signal 
efficiency becomes too low despite the formally better sensitivity.  

\subsection{Reach in the $cc\met$ channel}\label{sec:charm}

This is the simplest, and in some sense, most model-independent
situation, which arises whenever the stop is the next-to-lightest
supersymmetric particle. Then, the only
two-body stop decay still open is
$\tilde t\rightarrow c \tilde\chi^0_1$ \cite{HK,BDGGT,BST}.
In the absence of any flavor-changing effects in the squark sector
(which is rather unlikely), this decay proceeds through a loop,
otherwise it occurs at tree-level through stop-scharm mixing.
The stop signature in this channel is two acolinear 
charm jets and missing transverse energy carried away by neutralinos.
The events for this analysis in Run I were collected using a trigger 
which required $\met > 35$ GeV.
We select events with 2 or 3 jets with $E_T > 15$ GeV and 
$|\eta| < 2$.  The $\met$ cut is increased beyond the
trigger threshold to 40 GeV and we 
require that the $\met$ is neither
parallel nor anti-parallel to any of the jets
in the event in order to reduce the contribution
from the processes where the missing
energy comes from jet energy mismeasurement: 
$\min\ \Delta\varphi(\met,j) > 45^\circ$,
$\Delta\varphi(\met,j_1) < 165^\circ$, 
and $45^\circ<\Delta\varphi(j_1,j_2)<165^\circ$,
where the jets are ordered in $E_{T}$. We also veto electrons and muons to 
suppress the W+jets background.

We use the JP algorithm to tag a charm jet requiring that at least one 
jet has a probability less than 0.05.
This requirement, chosen to optimize the expected signal significance, rejects
97\% of the background while its efficiency for the signal is 25\%.

As can be seen from Table~\ref{backgrounds},
the dominant source of remaining background for this analysis is $W$+jets 
production where the vector boson gives a hadronically decaying $\tau$ lepton. 
There is also a contribution from QCD multijet production. The middle plot 
in Figure~\ref{eff} shows the signal efficiency as a function of the
stop and neutralino mass difference.
\begin{figure}[t]
\centerline{\psfig{figure=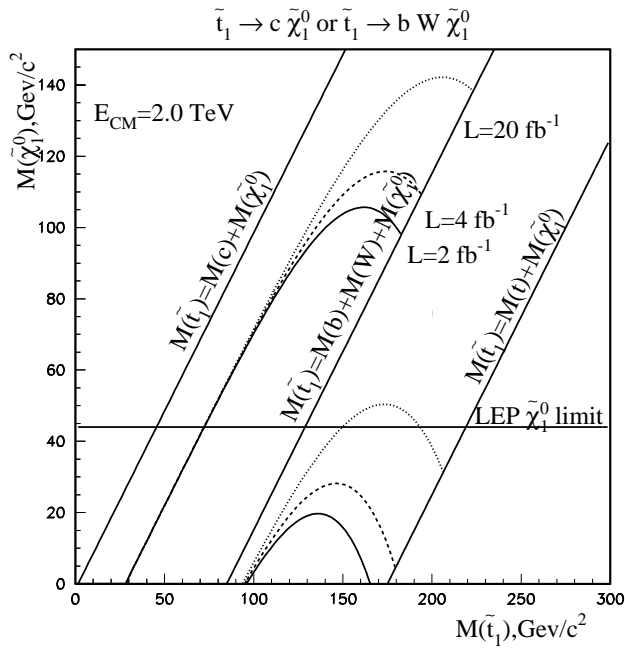,height=3.5in}}
\begin{center}
\parbox{5.5in}{
\caption[] {\small The same as Fig.~\ref{sens_bchar},
for the light stop search in the
$cc\met$ and $bj\ell\met$ channels.
The $cc\met$ ($bj\ell\met$) analysis assumes 100\%
branching ratio of $\tilde t\rightarrow c \tilde\chi^0_1$
($\tilde t\rightarrow b W^+ \tilde\chi^0_1$).
\label{sens_2or3}}}
\end{center}
\end{figure}

Figure~\ref{sens_2or3} shows the Run~II exclusion contours
in this channel for several integrated luminosities.
CDF Run~Ib results for this channel were presented in \cite{Ibcc}.
The reach in stop mass is determined by the accumulated 
statistics while the reach in neutralino mass depends on the efficiency of our 
selection cuts, where the most limiting is the $\met$ cut, effectively
fixed by the $\met$ trigger threshold. This is what determines the gap
between the kinematic limit and the excluded region. 
In Run~II CDF will have a possibility to use the secondary vertex information 
at the trigger level \cite{SVT}. Addition of a displaced track requirement 
may allow to lower the $\met$ trigger threshold to 25 GeV and, therefore, 
significantly extend the excluded region to the kinematic limit.
Even with the high $\met$ cut, we see that already with 2~fb$^{-1}$
the Tevatron will be able to probe regions well beyond
the sensitivity of LEP searches.

If the mass gap between the stop and neutralino masses is
larger than $m_b+M_W$, the three-body decay 
$\tilde t\rightarrow b W^+ \tilde\chi^0_1$
opens up. We use the signature $b\ell j \met$ 
discussed in Section~\ref{sec:chargino}
to estimate the sensitivity for this kinematic region 
(see Fig.~\ref{sens_2or3}).

\subsection{Reach in the $jl^+l^-\met$ channel}
\label{sec:sneutrino}

If the stop is lighter than the chargino, but heavier than
any of the sleptons, then the three body decay modes
$\tilde t\rightarrow b \ell^+ \tilde \nu_\ell$ and
$\tilde t\rightarrow b \tilde \ell^+ \nu_\ell$ become dominant
\cite{HK,BDGGT,3body}. 
Such a situation may readily arise in the minimal SUGRA model.
As we mentioned in Sec.~\ref{sec:chargino}, for $M_0\le 0.55 \cdot M_{1/2}$,
the sneutrinos are lighter than the chargino. In those cases, there
almost always exist values for $A_0$ (both negative and positive)
which will bring the light stop mass in between
$m_{\tilde \nu}$ and $m_{\tilde\chi_1^+}$.

\begin{figure}[t]
\epsfysize=3.5in
\epsffile[-120 165 180 635]{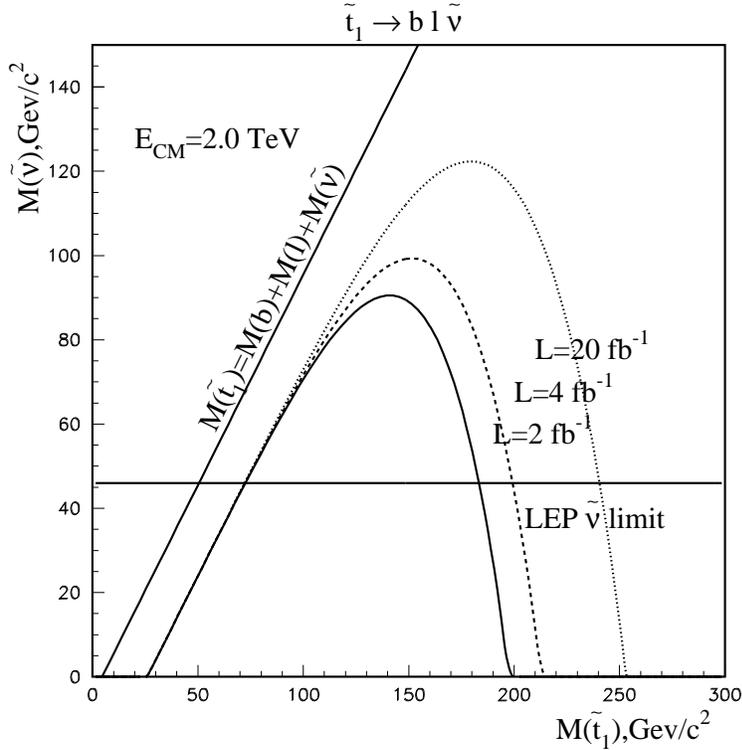}
\begin{center}
\parbox{5.5in}{
\caption[]
{\small The same as Fig.~\ref{sens_bchar}, but for
the light stop search in the $b\ell^+\ell^- j \met$ channel,
assuming that the stop decays as 
$\tilde t\rightarrow b \ell^+ \tilde \nu_\ell$ and
$\tilde t\rightarrow b \tilde \ell^+ \nu_\ell$.
\label{sens_blsn}}}
\end{center}
\end{figure}

Since in this case the stop leptonic branching ratio is high, we can
afford to require two leptons in the final state. A $b$-tag
is not required in order to save on jet acceptance.
We select events with two leptons $P_{T}(\ell_1)>8$ GeV/c
and $P_{T}(\ell_2)>5$ GeV/c, $\met>30$ GeV and at least one jet with 
$E_T>15$ GeV. In order to suppress the $b\overline{b}$
background, we require the leptons to be isolated --
the calorimeter energy sum in a cone of 0.4 in $\eta-\phi$ space around both
leptons should be less than 5 GeV.
Further cuts on the angle between either of the leptons and the
missing transverse energy reduce the background from jet mismeasurement:
$\Delta \varphi(\ell,\met)>20^o$ and
$\Delta \varphi({\rm dilepton\ system},\met)>20^o$.
The main backgrounds are Drell-Yan dilepton
production, top, $b\bar{b}$ and QCD multijet production 
(see Table~\ref{backgrounds}). The right plot 
in Figure~\ref{eff} shows the signal efficiency as a function of
the stop and sneutrino mass difference.
We show the reach of Run II in this channel in Figure~\ref{sens_blsn}.
Again, we see that the Tevatron will be able to go well beyond the LEP limits.

At large values of $\tan\beta$, the same two effects that can lead to
a light stop -- the enhanced RGE renormalization and the larger mixing,
can also make the lightest tau slepton
significantly lighter than the electron or muon sneutrino/slepton.
(Note that because of the large stau mixing, now the chargino can couple to
the lightest stau, so we are not limited to considering only the sneutrinos,
which are typically heavier.)
Therefore, there are larger regions in the mSUGRA parameter space where
one encounters the hierarchy
$m_{\tilde\tau} < m_{\tilde \chi_1^+}$ instead.
In the minimal SUGRA model this is possible for $M_0\lsim 5 M_{1/2}$.
In that case, the three-body stop decay 
$\tilde t\rightarrow b \tilde \tau^+_1 \overline{\nu}_\tau$
may be dominant, giving rise to the signature
$b\overline{b}\tau^+\tau^-\met$.
Since the leptons from the tau decays are rather soft, it may be
advantageous to consider a signature where
we replace one of the leptons with an identified tau jet:
$b\ell\tau j \met$ \cite{taus}.

\section{Higgsino LSP}\label{sec:stop to higgsino}

So far we have been considering only minimal SUGRA models,
which assume universality of the scalars and gauginos at the
unification scale $\mgut$. Theoretically, however, this assumption
is not very well motivated. If gravitational interactions are the mediator
of supersymmetry breaking, then universality would naturally hold at the 
Planck scale $\mplanck$ instead, and may get modified by
whatever physics there is between $\mgut$ and $\mplanck$.
If there is grand unification, however, the GUT symmetry will
preserve universality multiplet by multiplet. Thus one may expect that
sparticles belonging to the same GUT representation still have identical
soft masses at the GUT scale; while sparticles
belonging to {\em different} GUT representations may have
different soft masses. Within the framework of a SUSY SU(5) GUT,
this implies that the universal scalar mass parameter $M_0$ is now
being replaced by 4 scalar mass inputs at the GUT scale:
$M_{10}$, the mass of the two up-type squarks,
the left-handed down squark and the right-handed selectron;
$M_5$, the mass of the right-handed down-type squark and the
left-handed slepton doublet; and $M_{H_1}$ ($M_{H_2}$) ---
the soft mass for the down-type (up-type) Higgs doublet.
This type of model has become known as the non-universal SUGRA model.

The non-universality in the boundary conditions for the scalar masses
affects the SUSY mass spectrum in several ways. First, and most
directly, it may change the
ratios of various scalar masses, for example the left-handed and
right-handed charged sleptons. However, the values of the scalar masses 
at low energies also depend on the gaugino masses through the RGE
evolution, and the effects from any scalar mass non-universalities
become diluted in the limit $M_0\ll M_{1/2}$.
A true test of universality will therefore require measuring several
squark or slepton masses to a very good precision, something which
may only be accomplished at the next linear collider.

There are also indirect implications of scalar mass non-universality.
The higgsino mass parameter $\mu$, which is determined from the
condition of electroweak symmetry breaking,
is sensitive to the soft mass spectrum
of the Higgses and third generation sfermions. In the minimal
SUGRA model, it turns out that typically the higgsino masses are
quite a bit larger than the gaugino ones, and as a result, the LSP
and the lightest chargino are mostly gaugino-like\footnote{ 
There is one possible exception to this rule -- for
multi-TeV $M_0$ one can find natural \cite{naturalness}
regions of $|\mu|<M_1,M_2$ \cite{small mu}, where $M_1$ and $M_2$
are the SUSY breaking mass parameters for the $U(1)_Y$ and $SU(2)_L$
gauginos, respectively.}. In the non-universal SUGRA model,
where the two Higgs soft masses are free inputs at the GUT scale,
we often find regions of parameter space where $|\mu|<M_1$, and
as a result, the two lightest neutralinos and the light chargino
are almost degenerate and mostly higgsino-like.
Relaxing the universality assumption for the gaugino masses
may also lead to higgsino-like LSP \cite{nonuniv gaugino}.

The implicit assumption for all three stop searches in the  
previous Section was that the LSP is gaugino-like. In case of
a higgsino-like LSP, the search strategy obviously 
needs to be modified.

The analysis in Sections~\ref{sec:chargino} relies on the
presence of a hard lepton from the chargino decay. If the chargino is
higgsino-like, it is very close in mass with the LSP and the leptons
from the chargino decays are too soft to be used for either triggering
or off-line. The analysis in Section~\ref{sec:sneutrino} assumes
the existence of a significant mass gap between the LSP and the
lightest chargino, bigger than the gap between the LSP and the
stop itself. This is not true if the LSP is higgsino-like.
In fact, the dominant stop decay in that case is
$\tilde t\rightarrow b \tilde\chi_1^+$. The subsequent
chargino decay to $\tilde\chi_1^0$ is associated with
very soft leptons or jets. The only observable signature therefore
is $b\overline{b}\met$ \cite{susy98}. It is similar to the signature
considered in Section~\ref{sec:charm}, except that now the heavy
flavor jets are b--jets. To understand our sensitivity to
stop in this channel we apply all the cuts discussed
in Section~\ref{sec:charm}. We gain
some sensitivity with respect to the $cc\met$ analysis
because the heavy flavor tagging technique that we use is more
efficient to bottom than to charm.

\begin{figure}[t]
\centerline{\psfig{figure=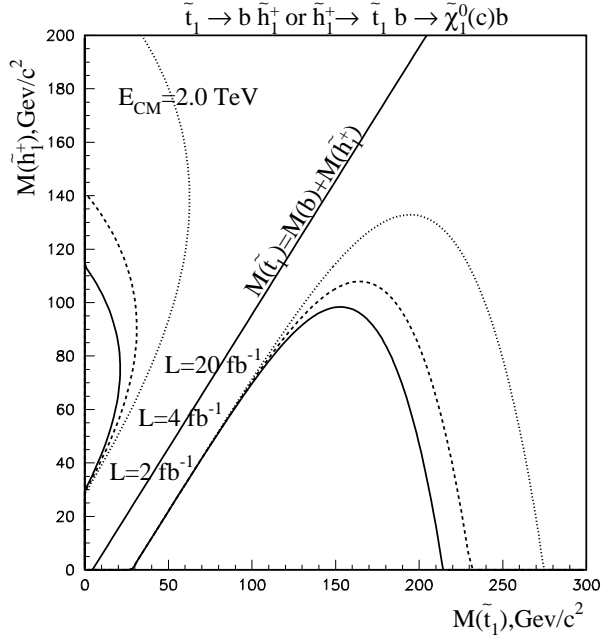,height=3.5in}}
\begin{center}
\parbox{5.5in}{
\caption[] {\small Sensitivity to $\tilde t\rightarrow b \tilde h^+$ or
$\tilde h^+ \rightarrow b \tilde t$ decays
for different integrated luminosities.
\label{massh}}}
\end{center}
\end{figure}

A family of curves below the diagonal in Fig.~\ref{massh}
represent the reach of Run II in the $b\overline{b}\met$
channel for different integrated luminosities.
We see that we restore our sensitivity to stop discovery even
though the higgsino decay products are lost. The reach is similar
to the one presented in Fig.~\ref{sens_2or3} for the $cc\met$ channel.

\section{Stop search in chargino (higgsino) production}
\label{sec:higgsino to stop}

It may be that the stop is light, yet cannot be found by regular
means since it is almost degenerate with the LSP and its decay
products are too soft to be observed. Such regions of the SUGRA
parameter space readily exist, and under those rather
unfortunate circumstances one should look for
stops among the decay products of some other, rather
light particles\footnote{In the case of an extreme degeneracy with
the LSP (which is somewhat preferred on the basis of relic
density arguments \cite{relic density}),
the stops can be long lived and form bound states, ``top squarkonia''
\cite{top squarkonium}, which eventually decay through
$\tilde t_1\tilde t_1^\ast$ annihilation.}. Let us review our options.

Of all the SM particles, the only ones likely to be heavier
than the stop are the top and possibly the Higgs(es),
but the production cross sections for the latter
at the Tevatron are too small to be relevant.
Top quarks may in principle decay to stops and gluinos
\cite{top to stop gluino}.
This channel is usually closed, since the gluino is typically
quite heavy. Some unconventional models \cite{heavy gluino}
predict a light gluino in the tens of GeV range,
with the gluino possibly being the LSP.
However, existing data already rule out
the range of gluino masses for which the two-body decay
$t\rightarrow \tilde t_1 \tilde g$ is open \cite{heavy gluino}.
Top quarks may also decay to stops and neutralinos
\cite{BDGGT,top to stop neutralino}.
One can look for these decays through precise
measurements of the top branching ratios.
If the stop is really degenerate with the LSP, it decays
invisibly, and as a result the signature is an invisible top.
If the $\tilde t-\tilde\chi_1^0$ mass difference is
large enough so that the $c$-jets can be detected, yet
small enough to evade the stop search in the $cc\met$ channel,
then the signature will be top quarks decaying to
$t\rightarrow c\met$. The top cross section is big,
but the width into $Wb$ is also quite large, so these
will be quite challenging analyses. 

We now turn to discuss the possibility of producing stops
in SUSY cascades. Among the remaining SUSY particles,
gluinos have the largest production cross section, 
and they can decay to $t\tilde t$ pairs
\cite{Bartl91,BDGGT,gluino to stop}.
Since the stops are invisible, the
signature is similar to the leptonic channels of
top pair production. The crucial difference from
$t\bar{t}$ production is that because of the Majorana nature of the gluino,
half of the time the top quarks will have the same sign.
Such an analysis is also in preparation for Run II.

One can also consider neutralino decays to top-stop. In this case, however,
the neutralinos would have to be heavier than $m_t+m_{\tilde t}$.
Since their cross sections are electroweak, they would be too small
to be observed at the Tevatron.

Stops may also appear in sbottom decays:
$\tilde b\rightarrow \tilde t W^-$ or
$\tilde b\rightarrow \tilde t H^-$, but first, these processes will
have to compete with $\tilde b\rightarrow b \tilde\chi^0_1$,
which is preferred by phase space, and second,
since stops are invisible, the final state signatures 
($W^+W^-$, $W^+H^-$ or $H^+H^-$) will have very large backgrounds.

Finally, we can consider production of charginos, which later decay
to stops: $\tilde \chi^+\rightarrow \tilde t\, \overline{b}$
\cite{susy98}.
This case looks more promising than the neutralino decays to stops.
First, there is much less phase space suppression,
and second, for gauginos, the chargino pair production cross sections
are larger than the neutralino ones. In the rest of this Section we
shall consider this channel in more detail.

We start by assuming that the LSP is mostly Bino. (If it were higgsino, we
are back to the case discussed in Section~\ref{sec:stop to higgsino}).
Then, if the SUGRA relations among the gaugino masses hold, the
gaugino-like chargino would have to be twice as heavy. We shall consider
$\mu$ as a free parameter, which can be easily
accounted for by non-universalities as discussed previously.
Then, if $|\mu|\ll M_2$, the lighter chargino will be higgsino-like;
while if $|\mu|\gg M_2$, we are back to the typical mSUGRA case
of gaugino-like chargino.

\begin{figure}[t]
\centerline{\psfig{figure=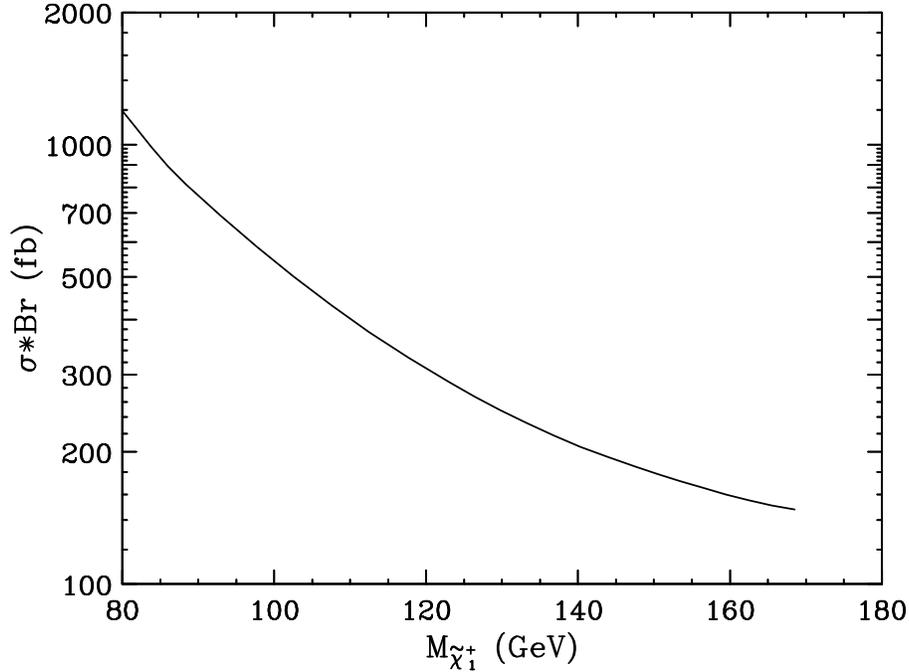,height=3.5in}}
\begin{center}
\parbox{5.5in}{
\caption[] {\small $\sigma\ast {\rm Br}\equiv
\sum_{i,j}\sigma(\tilde\chi^+_i\tilde\chi^-_j)
Br(\tilde\chi^+_i\rightarrow\tilde t_1 \bar{b})
Br(\tilde\chi^-_j\rightarrow\tilde t^\ast_1 b)$
as a function of the chargino mass
$m_{\tilde \chi^+_1}$, which is varied by changing $\mu$. 
We fix $M_2 \sim 2M_1 = 180$ GeV, and appropriately adjust $A_t$ 
so that the lightest stop is degenerate with the LSP.
The rest of the sparticle spectrum is taken to be very heavy.
\label{lstophiggsino}}}
\end{center}
\end{figure}

Since the stop is almost degenerate with the neutralino
by our assumption in this section,
the resultant charm jets from its decay are very soft and
cannot be detected, so the final state signature will be
$b\overline{b}\met$ and we can consider the same selection
which was used in Section~\ref{sec:charm}.
Yet the production cross section is different, 
since in this case we need to produce charginos via a weak process.
In Fig.~\ref{lstophiggsino}
we show a plot of the signal cross section times branching ratio
versus the lightest chargino mass $m_{\tilde \chi^+_1}$.
The mass, as well as the gaugino/higgsino mixture of the
lightest chargino in the figure are varied by changing $\mu$,
simultaneously adjusting $A_t$ so that to keep the lightest stop
degenerate with the LSP. The rest of the sparticle spectrum
is assumed to be very heavy. We have also included the contributions
from relevant processes with heavier neutralinos which may decay 
to $\tilde\chi_1$. We use conservatively leading order
results for chargino/neutralino production; next-to-leading
order QCD corrections can increase the signal by 20-30\%
\cite{ino xsections}. We see from Fig.~\ref{lstophiggsino}
that the signal cross section is rather small, and for the
range of chargino masses beyond the LEP coverage, the
signal is only a few hundred fb.

Our reach of Run II in this channel is presented by a 
family of curves above the diagonal in Fig.~\ref{massh}.
Again, we can recover some region in parameter
space by using an alternative signature in our search,
but the absolute reach is not very impressive,
mostly because of the small production cross sections.

\section{Light bottom squarks}\label{sec:sbottoms}

Light sbottoms in the mSUGRA model can appear only at
small $M_{1/2}$ {\em and} small $M_0$. In addition, they are
always accompanied by light stops as well. In fact, throughout the 
whole mSUGRA parameter space, $m_{\tilde b_1}> m_{\tilde t_1}$.
The only exception appears at values of $\tan\beta> 20$ and $\mu<0$,
where we find that $-60 \ {\rm GeV} < m_{\tilde b_1}-m_{\tilde t_1}<0$.
(The correlation with the sign of $\mu$ is due to the SUSY
threshold corrections to the bottom Yukawa coupling \cite{yb corr}.)
However, this part of parameter space is severely constrained
\cite{bsg on susy}
by the $b\rightarrow s\gamma$ measurement from LEP \cite{bsgLEP}. 
Similar conclusions hold even when we relax scalar mass
universality.

\begin{figure}[t]
\centerline{\psfig{figure=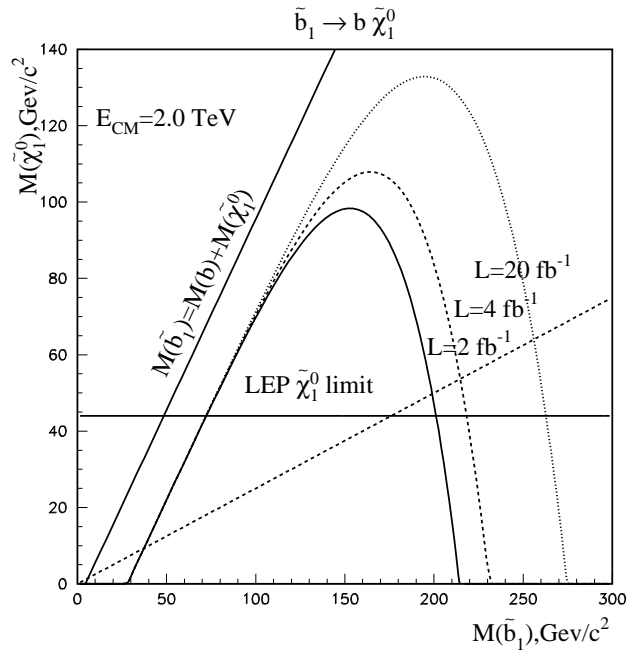,height=3.5in}}
\begin{center}
\parbox{5.5in}{
\caption[] {\small Sensitivity of the light sbottom search in the bottom
neutralino channel for different integrated luminosities.
\label{sens_sb}}}
\end{center}
\end{figure}

In any case, having 
done the analysis for stops decaying to higgsinos, it is
straightforward to extend it to the case of sbottom production and
sbottoms decaying directly to LSP's. Again we can consider the 
same selection used in Section~\ref{sec:charm} to estimate our
sensitivity to a direct sbottom search.
Our reach in Run II is presented  in Fig.~\ref{sens_sb}.
Although the cross section of sbottom production is roughly equal to the
stop production cross section for the same squark mass, our reach
in sbottom mass is somewhat higher than in stop mass in the similar
$c\bar{c}\met$ channel -- compare 
to Fig.~\ref{sens_2or3}. This is due to the higher bottom
tagging efficiency as compared to charm tagging efficiency.

By scanning the mSUGRA parameter space, we
found that the universality assumption leads to the relation
$m_{\tilde\chi^0_1}\le 0.25\ m_{\tilde b_1}$ so that the mSUGRA
parameter space maps onto the region below the dotted line
in Fig.~\ref{sens_sb}.
Relaxing scalar mass universality, we still find that typically
$m_{\tilde\chi^0_1}/m_{\tilde b_1}\lsim 0.25$,
but this ratio may go up to 0.4 for $\tan\beta>20$ and $\mu<0$,
which is in conflict with $b\rightarrow s\gamma$.
This means that the observation of a signal in this channel
already in Run II, if interpreted as sbottom production,
will hint towards a more unconventional low-energy SUSY,
for example non-universal gaugino masses.

\section{Stop (sbottom) as NLSP in gauge-mediated models}\label{sec:GM}

Gauge mediation is an intriguing alternative for communicating
supersymmetry breaking to the visible sector (MSSM) \cite{GM}.
It offers the potential of solving the supersymmetric flavor problem,
and leads to novel collider phenomenology \cite{GMpheno}.

The minimal gauge-mediated models, where the only SUSY
breaking contributions to the scalar masses are from SM gauge loops,
do not predict a light stop in the spectrum. However, one can easily 
imagine non-minimal extensions with extra gauge groups
\cite{GMgaugegroups} or more complicated messenger sectors \cite{GMmess},
which may lead to a light stop. Gauge mediated models are characterized by
a Goldstino LSP $\tilde G$, which is almost massless.
Of course, all our previous
results hold for the case of gauge mediated models with a stable
neutralino NLSP, since in that case the phenomenology is no different
from SUGRA models. But what is more, our results from
Sections \ref{sec:charm} and Section~\ref{sec:sbottoms} may be applied
for the case of stop and sbottom NLSP, which decays promptly to the
Goldstino. Identifying the neutralino LSP with the Goldstino, and 
taking the limit $m_{\tilde\chi^0_1}\rightarrow 0$, we can
read off the Run II stop (sbottom) mass reach 
from the $x$-axis in Fig.~\ref{sens_2or3} (Fig.~\ref{sens_sb}).
For example, if the dominant prompt decay of the stop is
the three-body mode $\tilde t_1\rightarrow b W^+ \tilde G$
\cite{CP}, we can see that already with $2\ {\rm fb}^{-1}$
the Tevatron will be sensitive to light stop masses up to 160 GeV.

Prompt stop decays in gauge-mediated models can be expected
only if the SUSY breaking scale is very low. Otherwise,
the stops first hadronize in supersymmetric sbaryon or mesino states
and then decay over macroscopic distances, leading to events with
highly-ionizing tracks or displaced jets and large $\met$
\cite{mesino}. It was recently pointed out, that in the case
of prompt decays, stop mesino-antimesino oscillations
can provide a very distinctive signature of like-sign top quark
events \cite{mesino}.

\section{Conclusions}\label{sec:conc}

There are various ways to look for a light stop, depending on
the rest of the sparticle spectrum. We presented a long
list of stop signatures which can be looked for in Run II
of the Tevatron with upgraded detectors. There are also
numerous other possible stop signals. For example,
models with broken R-parity
allow for a set of stop decays beyond the ones considered here
\cite{RPV decays}, as well as associated (single) stop production
processes \cite{single stop}. Also, the 4-body decay of the stop
can become dominant in certain models with suppressed
flavor-changing effects \cite{4-body}.

In most cases the upcoming Tevatron runs can
extend the LEP-II reach~\cite{LEP}.

\vspace{.25in}

{\it Acknowledgements:} We would like to thank M.~Carena
and M.~Peskin for discussions.
Fermilab is operated by URA under DOE contract DE-AC02-76CH03000.

\newpage

\vfill

\end{document}